**Ferroionic $CuCrP_2S_6$-Integrated Silicon Photonics for Neuromorphic Computing**

**Ghada Dushaq*, Senior Member, IEEE, Srinivasa R. Tamalampudi, and Mahmoud Rasras*, Senior Member, IEEE**

*Photonics Research Lab, Department of Electrical and Computer Engineering, New York University Abu Dhabi, Abu Dhabi, UAE*
*ghd1@nyu.edu
*mr5098@nyu.edu

## Abstract

Neuromorphic photonic computing calls for optical states that combine stimulus accumulation, spontaneous relaxation, and nonvolatile state retention without continuous power dissipation. Here, we demonstrate a ferroionic photonic platform by integrating layered van der Waals $CuCrP_2S_6$ (CCPS) onto silicon microring resonators (MRR) using a back-end integration process. Electrical pulsing produces unidirectional blue shifts in the optical resonance, consistent with field-driven redistribution of mobile $Cu^+$ ions and giving rise to a history-dependent optical response. The resulting optical state is strongly pulse dependent, exhibiting both volatile and nonvolatile components. A short 20-pulse excitation produces an initial ≈ 70 pm resonance shift that relaxes toward baseline within ≈ 60 s, with a volatile relaxation time constant of $\tau_V \approx 27$–29 s. In contrast, after 1000 pulses, an initial ≈ 0.27 nm resonance shift relaxes to a persistent ≈ 0.13 nm nonvolatile state that is retained for at least 24 h. Repeated +6 V electrical pulses produce a progressive resonance shift over 1000 cycles, demonstrating cumulative optical integration. A fixed-wavelength readout converts the accumulated resonance shift into a thresholded optical response, with the normalized transmission remaining above the selected threshold through 600 pulses and falling below it at the 800-pulse state. These dynamics realize the basis for a leaky integrate-and-fire (LIF) photonic neuron, where the pulse-dependent ionic response provides integration, spontaneous relaxation supplies a passive leak, persistent states provide nonvolatile memory, and fixed-wavelength readout provides an optical threshold response.

## I. Introduction

Photonic neural-network hardware offers high-bandwidth information processing and parallel implementation of linear operations[1], [2], [3]. However, scalable programmable photonic systems require efficient mechanisms for updating and retaining optical states within integrated photonic circuits[4], [5]. Conventional tuning approaches in silicon photonics involve distinct performance trade-offs. Thermo-optic phase shifters are technologically mature and provide large tuning ranges, but are volatile and require continuous electrical power to maintain a programmed state[6], [7]. Free-carrier-based tuning supports high-speed operation but can introduce additional optical loss through free-carrier absorption[8], [9]. Phase-change materials (PCMs), in contrast, provide nonvolatile optical states through thermally induced transitions between material phases and are well suited for persistent optical storage[10], [11]. However, conventional nonvolatile states are designed for persistent optical storage and do not inherently provide the spontaneous relaxation dynamics required for leaky temporal processing[10], [12]. For neuromorphic photonic systems, an important opportunity is to develop physical state variables that combine stimulus accumulation, spontaneous relaxation, and persistent state retention, thereby supporting neuronal and memory functions directly through device dynamics[13], [14].

Ferroionic van der Waals (vdW) materials provide an attractive route toward neuromorphic functionality because ionic transport occurs within a polar crystalline lattice, enabling electrically accessible and history-dependent material states[15], [16], [17], [18]. Among these materials, copper chromium thiophosphate ($CuCrP_2S_6$, CCPS) contains mobile $Cu^+$ ions that occupy off-centered sites within its layered structure and can redistribute under an applied electric field[19], [20]. The resulting ionic response depends on the amplitude and temporal characteristics of the electrical excitation and can modify electronic transport through field-driven $Cu^+$ redistribution and interfacial processes at the metal/CCPS contacts[21], [22]. In lateral CCPS devices, these dynamics are associated with electrical hysteresis and resistive switching between high- and low-resistance states. When CCPS is integrated with silicon photonic structures, the electrically driven evolution of its material state can modify the optical properties experienced by the guided mode, translating the ferroionic response into measurable shifts of the optical resonance[20].

In this work, we integrate mechanically exfoliated CCPS onto silicon microring resonators with coplanar lateral electrodes using a back-end transfer process[23], [24], [25]. Electrical pulsing produces unidirectional resonance shifts consistent with field-driven $Cu^+$ redistribution, giving rise to a history-dependent photonic ionic response. Following a short 20-pulse programming sequence, the optical response is predominantly volatile, whereas increasing the cumulative excitation to 1000 pulses at the same pulse amplitude and width produces a persistent nonvolatile component with retention demonstrated for at least 24 h. Under repeated +6 V pulses, the resonance progressively shifts over 1000 pulses, demonstrating cumulative optical integration. At a fixed readout wavelength, the corresponding normalized transmission remains above the selected optical threshold through 600 pulses and falls below it at the 800-pulse state. The pulse accumulation, spontaneous relaxation, nonvolatile retention, and optical thresholding provide the physical basis for a leaky integrate-and-fire (LIF)-like photonic neuron response within a single ferroionic–photonic device.

We investigate how these material dynamics translate into optical state evolution and neuromorphic functionality. Section II explains the integration and ferroionic response of CCPS photonic devices and their lateral electrical characteristics and unidirectional resonance shifts. Section III examines cumulative resonance evolution under repeated electrical excitation as an optical integration mechanism. Section IV quantifies the pulse-dependent coexistence of volatile relaxation and persistent state retention. Section V maps these experimentally observed dynamics onto a leaky integrate-and-fire (LIF)-like photonic response using fixed-wavelength optical readout and thresholding. Section VI discusses the implications and limitations of the demonstrated behavior for neuromorphic photonic systems, followed by the conclusions in Section VII.

## II. CCPS-Integrated Photonic Devices and Ferroionic Response

### A. Ferroionic CCPS and Photonic Integration

CCPS is a layered van der Waals (vdW) ferroionic material characterized by electrically driven redistribution of mobile Cu ions within its polar crystal lattice[19], [26]. In contrast to conventional ferroelectric switching, which primarily involves local ionic displacement and domain evolution, CCPS additionally supports Cu-ion migration among multiple nonequivalent sites through intralayer (in-plane) and interlayer (out-of-plane) pathways[27], [28], as illustrated in Fig. 1(a). These multiple accessible ionic configurations enable gradual and history-dependent state evolution under electrical excitation[17], [21] .

The integrated photonic device is shown in Fig. 1(b) and (c). The microring resonator (MRR) was fabricated on a silicon-on-insulator (SOI) platform comprising a 220-nm-thick silicon device layer on a 2-μm buried oxide (BOX) layer. The add-drop MRR consists of a 460-nm-wide waveguide with a ring radius of 45 μm and a 100-nm coupling gap between the bus waveguide and the resonator, designed to provide efficient evanescent coupling and a high extinction ratio at the through port. Coplanar Ti/Au electrodes (10/100 nm) were prefabricated adjacent to the resonator, with a total metal thickness lower than the silicon waveguide height. Mechanically exfoliated CCPS was subsequently dry-transferred across the electrode and resonator region through 200 nm opening in the upper oxide cladding using the deterministic transfer process demonstrated in our previous work[23], [24], [25]. This configuration enables the lateral electrodes to apply an electric field across the CCPS while maintaining direct optical overlap between the transferred material and the silicon waveguide.

The optical properties of the exfoliated CCPS were characterized by high-resolution spectroscopic ellipsometry. The wavelength-dependent refractive index (n) and extinction coefficient (κ) extracted from CCPS on a 300-nm $SiO_2$/Si substrate were used as material inputs for the optical mode analysis. CCPS has a reported bandgap of approximately 2.1 eV[20], [29], corresponding to an absorption edge well above the photon energy at telecommunication wavelengths (~0.8 eV at 1550 nm). Consequently, multilayer CCPS exhibits high transparency and low optical absorption in the short-wave infrared (SWIR) region[20], [25], [29]. Additionally, CCPS has a relatively high refractive index [30], therefore, these properties make CCPS well suited for integration with high-index-contrast silicon photonic waveguides, providing appreciable modal interaction while maintaining low material-induced optical loss.

Eigenmode simulations were performed using the MODE solver in the Ansys Lumerical Device Multiphysics Simulation Suite. As shown in Fig. 1(c), the silicon waveguide predominantly supports a quasi-TE mode that is strongly confined within the waveguide while extending evanescently into the transferred CCPS layer. For a multilayer CCPS flake with an average thickness of approximately 60 nm, the mode analysis at 1550 nm shows that approximately 5% of the guided optical power is confined within the CCPS layer. This modal overlap provides sufficient interaction between the guided field and the ferroionic material for electrically induced changes in the CCPS optical properties to modify the effective index experienced by the guided mode and, consequently, the MRR resonance. This back-end integration approach therefore introduces ferroionic functionality after fabrication of the underlying passive silicon photonic circuit.

**B. Electrical Programming and Mode-Dependent Ionic Response**

The nonlinear and hysteretic electrical response shown in Fig. 1(d) demonstrate the history-dependent transport behavior of the integrated CCPS device. Under a lateral electric field, mobile Cu ions migrate toward the corresponding electrode, consistent with established field-driven Cu-ion transport in CCPS[19], [31]. In the present device geometry, the guided optical mode is strongly confined around the silicon waveguide between the lateral electrodes [Fig. 1(e)]. Consequently, field-driven Cu-ion redistribution changes the ionic population within the optical mode-overlap region, modifying the effective refractive index, $n_{eff}$, experienced by the guided mode.

For electrical programming, voltage pulses were applied through the lateral electrodes while the optical response of the MRR was monitored using TE-polarized light. Light from a tunable laser (Keysight 8164B Lightwave Measurement System) was edge-coupled into the device using a lensed fiber, and the transmitted light was collected by an output lensed fiber and recorded using a power meter. The input polarization was calibrated using reference microring resonators fabricated on the same chip with identical geometries. Unless otherwise specified, the measurements reported here were performed using TE-polarized light near 1550 nm. Transmission spectra were recorded after electrical excitation, and the resonance displacement was defined as $\Delta\lambda=\lambda_r-\lambda_{r,0}$ where $\lambda_{r,0}$ and $\lambda_r$ denote the initial and programmed resonance wavelengths, respectively.

Figure 1(f) shows the corresponding optical response under electrical excitation of opposite polarities. Both positive and negative bias shift the microring resonance toward shorter wavelengths relative to the initial unbiased state. Although reversing the electric-field polarity reverses the direction of Cu-ion migration, the lateral device geometry drives ions toward either electrode and therefore away from the central region of strongest optical confinement [Fig. 1(e)]. We attribute the resulting same-sign blue shifts to this mode-dependent ionic redistribution. Migration toward either electrode reduces the Cu-ion population sampled by the guided mode, producing an effective-index change of the same sign. The magnitude of the resonance shift depends on the applied electrical stimulus and the preceding ionic state, establishing a history-dependent photonic response that forms the basis of the programmable dynamics investigated in the following sections. Unless otherwise specified, all measurements were performed at room temperature under ambient conditions.

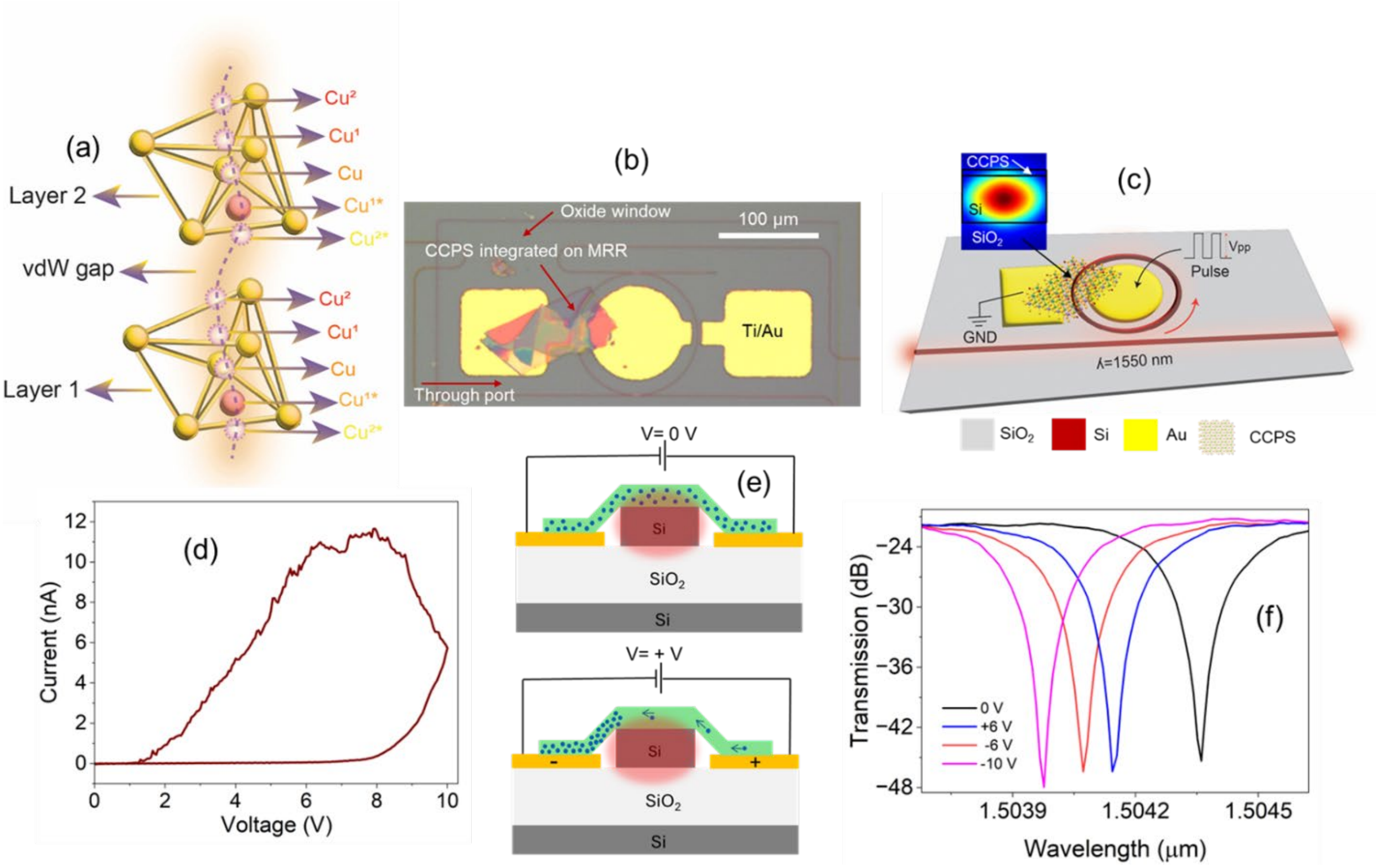

**Fig. 1. CCPS-integrated silicon photonic device and ferroionic response.** (a) Layered crystal structure of CCPS illustrating multiple Cu-ion sites and intralayer and interlayer migration pathways. (b) Optical micrograph of a mechanically exfoliated CCPS flake with thickness of ~ 55 nm integrated onto a silicon microring resonator (MRR) through an oxide-cladding window, with lateral Ti/Au electrodes. (c) Schematic of the electrically programmable CCPS-integrated MRR and calculated TE-mode profile showing the optical overlap with the CCPS layer. (d) Measured nonlinear and hysteretic current–voltage characteristic of the integrated CCPS device (measurements performed in dark). (e) Cross-sectional illustration of field-driven Cu-ion redistribution relative to the confined optical mode under zero and applied lateral bias. The redistribution modifies the Cu-ion population within the optical mode-overlap region. (f) Measured optical transmission spectra of microring resonator under positive and negative electrical bias, showing same-sign resonance blue shifts for opposite bias polarities.

## III. Cumulative Ferroionic Programming

Following the electrically induced state evolution established in Section II, we examined how the optical state evolves under repeated electrical excitation. Trains of +6 V pulses with a width of 250 ms and a repetition rate of 2 Hz were applied to the CCPS-integrated microring. Transmission spectra were recorded after cumulative pulse counts of 0, 200, 400, 600, 800, and 1000.

Figure 2(a) shows a progressive blue shift of the microring resonance with increasing pulse number. Successive pulse trains continuously displace the resonance in the same direction, demonstrating cumulative programming of the optical state. The corresponding resonance displacement, $\Delta\lambda$, extracted from each spectrum is plotted as a function of cumulative pulse number in Fig. 2(b).

The resonance shift increases progressively, reaching approximately 210 pm after 1000 pulses. Over the measured range, a linear fit gives an average accumulation of approximately 0.22 pm/pulse ($R^2 \approx 0.98$). The reduced incremental shift at higher pulse counts, however, indicates a tendency toward saturation. Thus, the measured response is not an ideal linear accumulation but a pulse-count-dependent evolution of the programmed optical state.

From a neuromorphic perspective, this cumulative response provides a physical analogue of integration, whereby successive electrical input pulses are mapped onto a progressively evolving optical state[3], [13]. Having established this cumulative programming behavior, we next examine how the programmed state evolves after removal of the electrical excitation, revealing the volatile and nonvolatile dynamics associated with leak and memory.

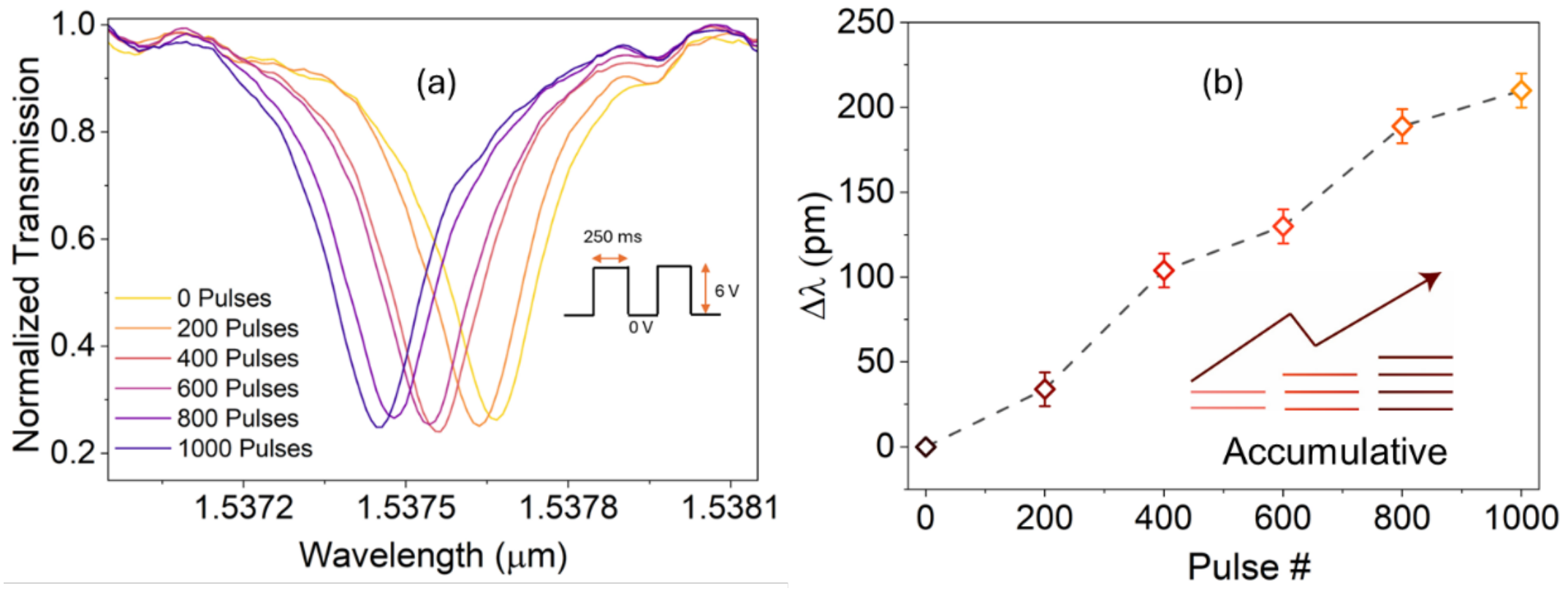

**Fig. 2. Cumulative ferroionic programming of the CCPS-integrated microring resonator.** (a) Measured normalized linear optical transmission spectra after 0, 200, 400, 600, 800, and 1000 cumulative +6 V pulses with 250 ms pulse width at 2 Hz. (b) Extracted resonance displacement Δλ as a function of cumulative pulse number, reaching approximately 210 pm after 1000 pulses. Error bars represent the uncertainty in determining the resonance wavelength.

## IV. Volatile and Nonvolatile Ferroionic Dynamics

### A. Pulse-Dependent Ferroionic State Evolution

The response of Cu-based ferroionic vdW materials depends on both the applied electric field and the temporal characteristics of the electrical excitation. Mobile Cu ions can access multiple nonequivalent configurations through intralayer and interlayer migration pathways, giving rise to electrical responses ranging from rapidly relaxing to persistent states[27], [32]. Figure 3(a) schematically illustrates this configurational freedom and the resulting binary, gradual, and accumulative state evolution.

Previous studies of Cu-based ferroionic materials have shown that the relative contributions of different ionic transport pathways can be modulated through the electrical programming conditions[32], [33]. As summarized conceptually in Fig. 3(b), lower-amplitude and shorter-duration excitation is associated predominantly with intralayer dynamics, whereas increasing the pulse amplitude or duration can promote coupled intralayer/interlayer transport and, under stronger excitation, a larger contribution from interlayer ionic migration. The regime boundaries in Fig. 3(b) are schematic and are included to illustrate the broader programming-dependent landscape of ferroionic dynamics.

In addition to pulse amplitude and duration, cumulative electrical excitation can influence the resulting ionic state. To examine this behavior experimentally in the CCPS-integrated microring, we compared the relaxation following two different cumulative pulse doses while maintaining the same pulse amplitude and width.

### B. Volatile Relaxation

A short programming sequence of 20 pulses (+6 V, 250 ms pulse width) produced an initial resonance blue shift of approximately 70 pm, as shown in Fig. 3(c). Following removal of the electrical excitation, the resonance progressively relaxed toward its initial state, with the shift decreasing to approximately 25 pm after 30 s and returning to approximately zero, within the experimental uncertainty, after about 60 s.

The relaxation can be described by an exponential decay,

$$\Delta\lambda_V(t)=\Delta\lambda_{V,0}\exp(-t/\tau_V) \quad (1)$$

where $\Delta\lambda_{V,0}$ is the initial volatile resonance displacement and $\tau_V$ is the characteristic relaxation time. Fitting the measured response yields $\tau_V \approx 27\text{–}29$s. The spontaneous recovery of the resonance after removal of the electrical input therefore provides a physical analogue of the leak process required in leaky temporal dynamics[34].

### C. Emergence of a Nonvolatile State

A higher cumulative programming dose produces a noticeably different relaxation response. Following 1000 pulses (+6 V, 250 ms pulse width, 2 Hz), the resonance exhibits an initial blue shift of approximately 270 pm [Fig. 3(d)]. After removal of the excitation, the shift decreases rapidly during the early relaxation period before approaching a residual value of approximately 130 pm. In contrast to the low-dose response, this residual displacement does not return to the initial state and remains stable, within the experimental uncertainty, throughout the 24-h measurement window.

To further examine the volatile contribution at higher cumulative excitation, the 1000-pulse relaxation was analyzed relative to the long-time nonvolatile plateau of approximately 130 pm. Fitting the relaxing component with an exponential decay yields an effective relaxation time of approximately 2.5 min (148 s), substantially longer than the 27–29 s relaxation observed following 20 pulses. This difference indicates that the volatile relaxation dynamics depend on the programmed ferroionic state rather than being characterized by a single, pulse-independent relaxation time.

The comparison between Fig. 3(c) and (d) therefore reveals a programming-dose-dependent change in the relative contributions of volatile and nonvolatile dynamics. Following the lower pulse dose, the programmed optical state is predominantly volatile and relaxes toward the initial resonance. Following the higher pulse dose, only part of the induced shift relaxes, while a substantial fraction persists as a nonvolatile optical state. This evolution is consistent with the progressive occupation of more persistent ferroionic configurations under cumulative electrical excitation.

The measured resonance displacement can therefore be represented as

$$\Delta\lambda(t)=\Delta\lambda_V(t)+\Delta\lambda_{NV}(t) \quad (2)$$

where $\Delta\lambda_V(t)$ represents the spontaneously relaxing component and $\Delta\lambda_{NV}(t)$ represents the persistent component over the measured time window. The coexistence of these two contributions enables distinct temporal functions within the same ferroionic photonic device. The volatile component provides a natural leak process, whereas the nonvolatile component provides optical state retention.

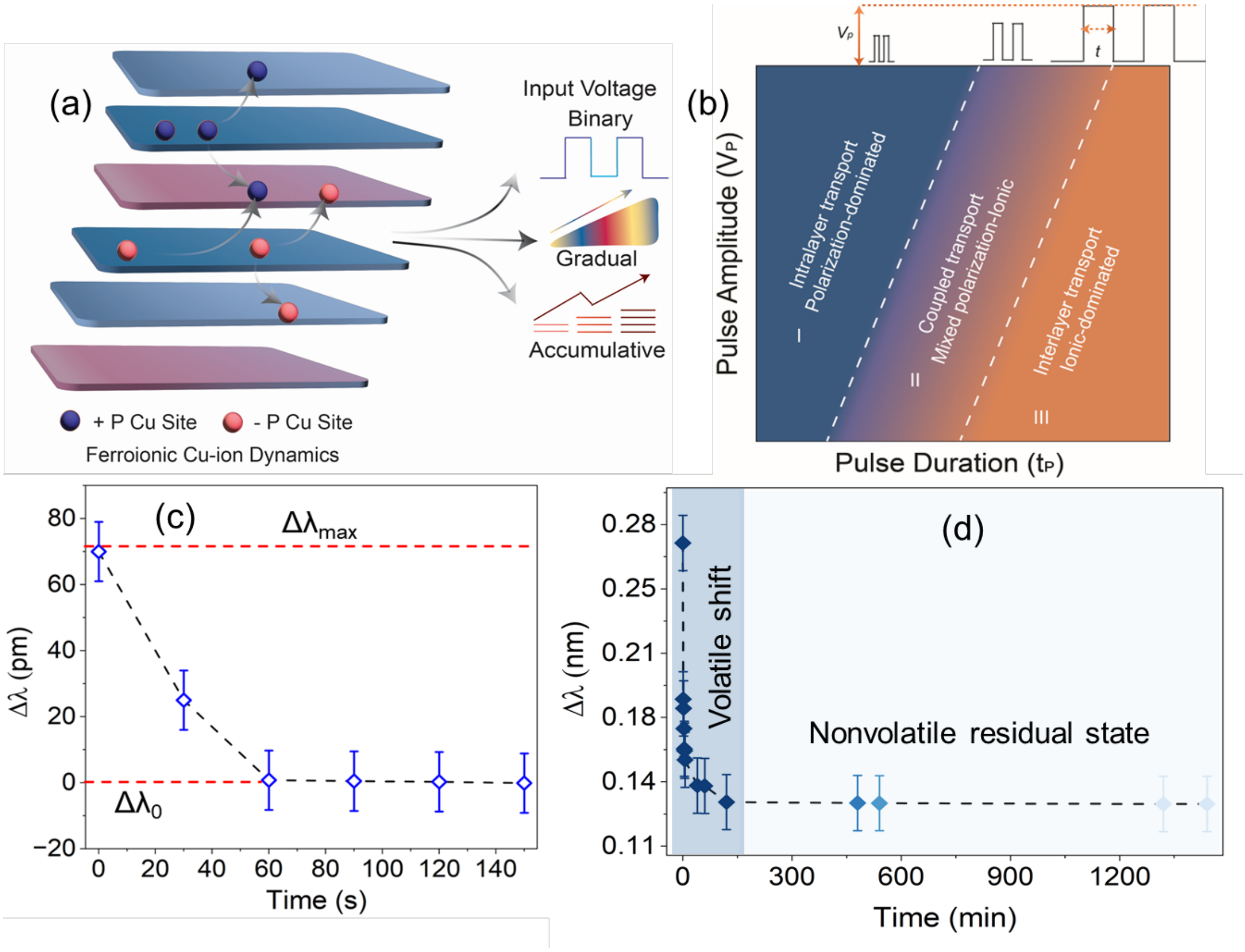

**Fig. 3. Programming-dependent volatile and nonvolatile dynamics in ferroionic photonics.** (a) Schematic illustration of mobile Cu-ion configurational dynamics in ferroionic vdW materials and the resulting binary, gradual, and accumulative responses to electrical excitation. (b) Conceptual map of pulse-dependent ferroionic transport as a function of pulse amplitude and duration, ranging from predominantly intralayer dynamics (I), through coupled intralayer/interlayer dynamics (II), to interlayer-dominated transport (III). (c) Measured relaxation following 20 +6 V pulses with a pulse width of 250 ms. (d) Measured relaxation and retention following 1000 +6 V pulses with a pulse width of 250 ms at 2 Hz.

## V. Leaky Integrate-and-Fire (LIF) Photonic Response

### A. Mapping the Ferroionic Dynamics to a LIF Response

The measurements in Sections III and IV establish the main physical processes required for a leaky integrate-and-fire (LIF) response. Repeated electrical pulses progressively shift the microring resonance, providing an optical state variable that accumulates with successive inputs. After the excitation is removed, the volatile component of this state spontaneously relaxes with a characteristic time constant of approximately 27–29 s, providing the leak process. At higher programming doses, a persistent resonance displacement remains, introducing a nonvolatile contribution to the optical state.

The measured resonance displacement can therefore be viewed as a state variable whose evolution is determined by the competition between pulse-driven accumulation and spontaneous relaxation. For the volatile component, this behavior can be represented as[34]

$$\frac{d\Delta\lambda_v(t)}{dt} = \frac{\Delta\lambda_v(t)}{\tau_v} + f[u(t), \Delta\lambda] \ (3),$$

where $\tau_v$ is the measured volatile relaxation time, $u(t)$ represents the electrical pulse input, and $f[u(t), \Delta\lambda]$ describes the experimentally observed pulse-dependent accumulation. The total resonance state additionally contains the persistent contribution $\Delta\lambda_{NV}$, as introduced in (2). This description is not intended as a microscopic model of the ionic transport, but rather as a mapping of the measured device dynamics onto a LIF-like state evolution.

### B. Optical Threshold Response

To translate the accumulated resonance displacement into an optical decision variable, the measured spectra were evaluated at a fixed readout wavelength ($\lambda_L$) positioned on the blue side of the initial resonance. As illustrated in Fig. 4(a), a readout wavelength of $\lambda_L$=1.50045 µm was selected, corresponding to a resonance-detuning threshold relative to the initial resonance ($\lambda_{r,0}$) of $\Delta\lambda_{th}=|\lambda_L-\lambda_{r,0}|$= (200±10) pm. As successive programming pulses shift the resonance toward $\lambda_L$, the transmission measured at this fixed wavelength progressively decreases. Figure 4(b) shows the measured optical transmission as a function of cumulative pulse number. The normalized linear transmission decreases from approximately 0.72 in the initial state to approximately 0.10 after 1000 pulses.

For the present demonstration, a normalized transmission level of $T_{th}$=0.2 was selected as the optical decision threshold. The transmission remains above this level through the 600-pulse state and falls below it at 800 pulses. Therefore, 800 pulses represent the first experimentally sampled state satisfying the threshold condition.

These results map the CCPS-integrated microring onto the principal elements of a LIF-like photonic response: successive electrical inputs accumulate in the resonance state, the volatile contribution provides spontaneous leak, and resonance detuning converts the evolving state into a thresholded optical output. The persistent component observed at higher cumulative excitation additionally provides nonvolatile state retention within the same device.

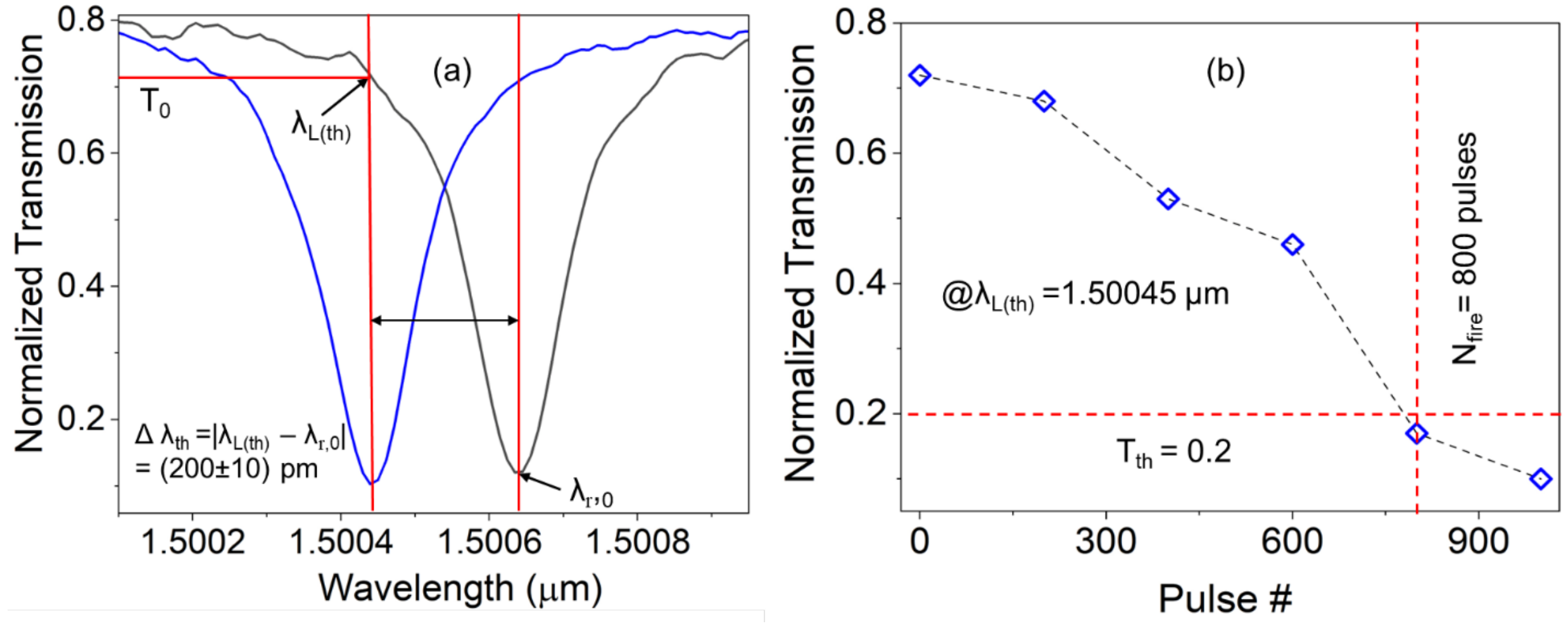

**Fig. 4. Optical transmission threshold response of the CCPS-integrated microring.** (a) Fixed readout wavelength $\lambda_L$=1.50045 positioned on the blue side of the initial microring resonance (slope of resonance). Cumulative electrical programming shifts the resonance toward the readout wavelength. (b) Normalized transmission at $\lambda_L$ as a function of cumulative pulse number.

## VI. Discussion

The results demonstrate that the optical response of the CCPS-integrated microring is governed by the history and temporal evolution of the electrically programmed ferroionic state. Repeated electrical excitation progressively shifts the resonance, whereas removal of the excitation reveals both spontaneously relaxing and persistent components. The relative contribution of these components depends on the programming conditions, allowing accumulation, relaxation, and state retention to be accessed within the same material device system.

The coexistence of volatile and nonvolatile dynamics is particularly relevant for neuromorphic photonics[35], [36]. The volatile component provides an intrinsic temporal decay that can represent leak, while the persistent component retains part of the programmed optical state without continuous electrical bias. Rather than requiring separate materials or devices for these functions, both responses arise from the relaxation dynamics of the electrically programmed CCPS state. This behavior provides a potential route for combining short-term temporal processing and persistent optical-state storage within a common photonic element.

The microring resonance also provides a means of converting the evolving ferroionic state into a threshold-based optical response. As shown in Fig. 4, cumulative resonance displacement produces a pronounced transmission change at a fixed readout wavelength. The optical decision point can therefore be selected through the readout wavelength and transmission threshold, without requiring an additional material transition to define the threshold. The measured accumulation, spontaneous relaxation, persistent state retention, and fixed-wavelength readout provide the device-level elements for LIF-like photonic operation.

The present experiments also identify aspects that remain to be developed toward complete neuronal operation. Further optimization of the electrode geometry, CCPS thickness, and optical-mode overlap could also be explored to reduce the programming voltage and pulse duration and to improve state reproducibility.

## VII. Conclusion

We demonstrated a ferroionic CCPS-integrated silicon microring who's electrically programmed optical state exhibits cumulative, volatile, and nonvolatile dynamics relevant to neuromorphic photonics. Repeated electrical pulses progressively shift the microring resonance, while removal of the excitation reveals a pulse dose-dependent balance between spontaneous relaxation and persistent state retention. The volatile response exhibits a characteristic relaxation time of approximately 27–29 s, whereas the programmed nonvolatile component remains stable within experimental uncertainty for at least 24 h. Fixed-wavelength readout further converts the accumulated resonance displacement into a thresholded optical transmission response. These results demonstrate that integration, leak, memory, and optical thresholding can be supported by the evolving ferroionic state of a single CCPS-integrated photonic device, providing a foundation for compact neuromorphic photonic systems.